\RequirePackage{lineno}
\documentclass[11pt]{article}
\usepackage{graphicx}
\usepackage{graphicx,amsmath,amsfonts}
\usepackage{longtable}
\usepackage[tight]{subfigure}
\usepackage{enumerate}

\newcommand{\BABARPubYear}    {08}

\newcommand{\BABARConfNumber} {004}
\newcommand{\SLACPubNumber} {13360}

\input babarsym

\def\BaBar{\babar\xspace}

\def\psip{\psitwos}
\def\smallISR{\scalebox{0.5}{ISR}}
\def\gISR{\ensuremath{\gamma_{\smallISR}}\xspace}

\long\def\inst#1{\par\nobreak\kern 4pt\nobreak
    {\it #1}\par\vskip 10pt plus 3pt minus 3pt}

\begin{document}
{\pagestyle{empty}

\begin{flushright}
\babar-CONF-\BABARPubYear/\BABARConfNumber \\
SLAC-PUB-\SLACPubNumber \\
\end{flushright}

\par\vskip 5cm

\begin{center}
Study of the $\pipi\jpsi$ Mass Spectrum \\
via Initial-State Radiation at \BaBar
\end{center}
\bigskip

\begin{center}
\begin{center}
\small

The \babar\ Collaboration,
\bigskip

%% author list as of 02-Jul-2008 (523 authors)
%
B.~Aubert,
M.~Bona,
Y.~Karyotakis,
J.~P.~Lees,
V.~Poireau,
E.~Prencipe,
X.~Prudent,
V.~Tisserand
\inst{Laboratoire de Physique des Particules, IN2P3/CNRS et Universit\'e de Savoie, F-74941 Annecy-Le-Vieux, France }
J.~Garra~Tico,
E.~Grauges
\inst{Universitat de Barcelona, Facultat de Fisica, Departament ECM, E-08028 Barcelona, Spain }
% ...

\end{center}\newpage

\mbox{ }\\
\today
\end{center}
\bigskip \bigskip

\begin{center}
\large \bf Abstract
\end{center}

We present an update of the study of the Y(4260) resonance, 
produced in the process $e^+e^-\to\gamma_{ISR}~\pipi\jpsi$ using initial-state 
radiation events at the PEP-II $e^+e^-$ storage rings. 
This study is based on 454\invfb of data recorded 
with the \BaBar detector at a center-of-mass energy in the 
$\Upsilon(4S)$ resonance region. 
From a fit with a single non-relativistic Breit-Wigner shape we  obtain updated parameters for the Y(4260) resonance which are
$m_Y = 4252 \pm6~^{+2}_{-3}$\mevcc and $\Gamma_Y = 105 \pm 18^{+4}_{-6}$\mevcc; we also measure
${\cal B}(\pipi\jpsi)\Gamma_{e^+e^-} = (7.5\pm 0.9~\pm0.8)$~eV.
We cannot confirm the recent BELLE observation of a broad structure
around 4.05\gevcc in this decay mode.

\vfill
\begin{center}

Submitted to the 34$^{\rm th}$ International Conference on High-Energy Physics, ICHEP 08,\\
29 July---5 August 2008, Philadelphia, Pennsylvania.

\end{center}

\vspace{1.0cm}
\begin{center}
{\em Stanford Linear Accelerator Center, Stanford University, 
Stanford, CA 94309} \\ \vspace{0.1cm}\hrule\vspace{0.1cm}
Work supported in part by Department of Energy contract DE-AC02-76SF00515.
\end{center}

\newpage
}

\section{INTRODUCTION}
\label{sec:Introduction}
The observation of X(3872) \cite{ct:X3872}, followed by 
the discovery of many other states such as the Z(3930) \cite{ct:Z3930}, the Y(3940) \cite{ct:Y3940} 
and the X(3940) \cite{ct:X3940}, has reopened interest in charmonium spectroscopy.
Some of these resonances cannot be fully explained by a simple 
charmonia model \cite{ct:charmonia}; four-quark state \cite{ct:tetra} and $D^0\overline{D}^{0*}$ molecule \cite{ct:molecule}
are some of the interpretations that have been proposed to explain their nature.
Among these new states the Y(4260) \cite{ct:babar-Y} that has been observed by \BaBar 
in the process $e^+e^- \to \gamma_{ISR}$ Y(4260)$ \to  \gamma_{ISR}~ J/\psi \pi^+\pi^-$, where $ISR$ denotes
initial state radiation. Being formed directly 
in $e^+e^-$ annihilation, it is known to have $J^{PC}=1^{--}$.
Nonetheless its properties do not fit with any 
simple charmonium interpretation and its nature is still unclear.
A recent ISR analysis by BELLE \cite{:2007sj} suggests the existence of a second state in the same decay mode, 
called Y(4008) \cite{:xiang}, with the same production mechanism.

After the discovery of Y(4260), other $J^{PC}=1^{--}$ states, produced through 
the same $e^+e^-$ initial state radiation mechanism, have been observed in 
the $\psi(2S) \pi^+\pi^-$ final state, such as the Y(4320) and the Y(4660)
\cite{Aubert:2006ge,:2007ea}.

In this letter, we present udated measurements of the resonance parameters and 
${\cal B}(\pipi\jpsi)\Gamma_{e^+e^-}$ of the Y(4260) based on the full \BaBar data set
which is approximately two times larger than the one used in the original analysis \cite{ct:babar-Y}.

\section{THE \babar\ DETECTOR AND DATASET}
\label{sec:babar}
The original \BaBar analysis \cite{ct:babar-Y} is updated here with the same method on a data set corresponding to a 
total integrated luminosity of 454\invfb obtained at the
SLAC PEP-II B-factory, running at the center-of-mass energy near the $\Upsilon$(4S) 
resonance.

The \BaBar detector is described in 
detail elsewhere \cite{ct:babar-detector}.  
Charged-particle momenta are measured in a tracking system consisting
of a five-layer double-sided silicon vertex tracker (SVT) and a
40-layer central drift chamber (DCH), both situated in a 1.5-T axial
magnetic field.
An internally reflecting ring-imaging Cherenkov detector (DIRC)
provides charged-particle identification together with $dE/dx$ measurements 
from SVT and DCH. A CsI(Tl) electromagnetic
calorimeter (EMC) is used to detect and identify photons and
electrons.
Muons are identified using information from the Instrumented
Flux Return (IFR) system, together with $E/p$, where the energy $E$ is
determined by the EMC and the momentum $p$ by the SVT and DCH.

\section{ANALYSIS METHOD}
\label{sec:Analysis}
In this analysis we reconstruct events:
\begin{equation}
e^+e^- \to \gISR \pipi\jpsi ,
\end{equation}
where $\gISR$ represents a photon that is radiated from the 
initial state $e^{\pm}$, lowering the center-of-mass energy of the $e^+e^-$ system. 
This mechanism allows to perform an energy scan of $e^+e^-$ interactions at the 
B-factories, where the center-of-mass energy of the interactions is 
fixed in the region of the $\Upsilon(4S)$ resonance.

A candidate $\jpsi$ meson is reconstructed via its decay to $e^+e^-$ or $\mu^+\mu^-$.
The lepton tracks, at least one of which must be identified as an electron or muon candidate, 
must be well reconstructed and must originate from the same vertex: a geometric 
fit of the $\jpsi$ candidate is conducted with beam-spot constraint. An algorithm to recover energy lost 
to bremsstrahlung is applied to electron candidates.
An $e^+e^-$ pair with its invariant
mass within the interval ($-$75,+40)~\mevcc around the PDG \cite{ct:PDG} $\jpsi$ mass
is taken as a $\jpsi$ candidate.
For a $\mu^+\mu^-$ pair, the interval is ($-$40,+40)~\mevcc.
The $\jpsi$ candidate is then kinematically constrained to
the nominal $\jpsi$  mass
and combined with a pair of oppositely-charged tracks
identified as pion candidates.

In the event reconstruction, we do not require observation of the 
initial state radiation photon $\gISR$,
as it is preferentially produced along 
the beam directions and it is not detected.

We select $e^+e^-\to\gISR\pipi\jpsi$ events with the following
criteria: the invariant mass squared recoiling against the $\pipi\jpsi$ system ($m_{rec}^2$) 
      is required to be in the range ($-$0.50,+0.75)~(\gevcc)$^2$;
the total number of reconstructed tracks in the event $\le$ 5.
Since the transverse component of the missing momentum ($p^{*}_{T,miss}$)
      is small for events with initial state radiation, we select events
      with $p^{*}_{T,miss} < 2.25$\gevc.
      The cosine of the angle between the $\ell^+$ momentum in the $\jpsi$ rest frame 
      and the $\jpsi$ momentum in the center-of-mass frame ($\cos\theta_l$) 
      is required to be $| \cos\theta_l | < 0.925$.

\begin{figure}[htbp]
\centering
    \includegraphics[height=7cm]{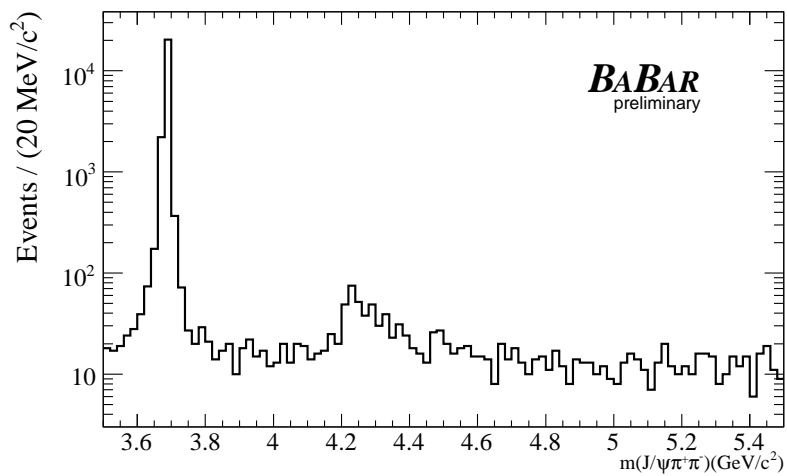}
   \caption{The invariant mass distribution of $\pipi\jpsi$ candidates with a logarithmic
   vertical scale.}
\label{fig:YMass_log}
\end{figure}

After this selection, a clean $\psip$ signal is apparent in 
the $\pipi\jpsi$ invariant mass spectrum, as shown in figure \ref{fig:YMass_log};
in the same distribution, a clean Y(4260) peak is also visible.

By means of Monte Carlo simulations we determine that the selection efficiency 
at the Y(4260) energy is 15.3\%.
The method used for the efficiency determination is checked at the $\psip$, 
where we obtain a value of the cross section that is consistent with the measurement performed by previous 
experiments \cite{:2007sj}. The $\psip$ mass is measured as $3685.35\pm0.02$\mevcc (where the error is statistical only);
the mass shift with respect to the PDG value \cite{ct:PDG} is taken into account as a systematic error in the Y(4260) mass
measurement.

The Monte Carlo $\pipi\jpsi$ invariant mass resolution and mass scale have been calibrated by comparing the widths of $\pipi\jpsi$ invariant 
mass distributions from $\psip$ decays in data and Monte Carlo.
We find that the Monte Carlo simulation
reproduces the observed r.m.s. of the $\pipi\jpsi$ distribution for the $\psip$ state.
The mass resolution is around 5\mevcc in the mass range $4.16\gevcc < m(\pipi\jpsi) < 4.36\gevcc$

\begin{figure}[tbp]
\centering
    \includegraphics[height=7cm]{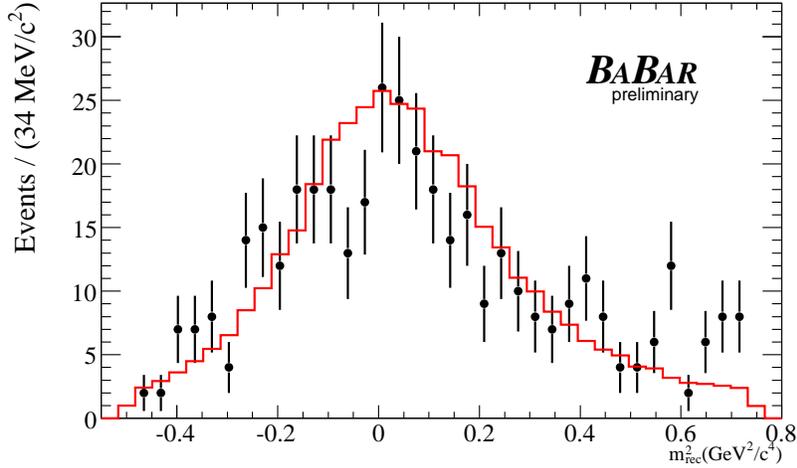}
   \caption{The distribution of $m_{rec}^2$. 
   The points represent the
data events passing all selection criteria
and having a $\pipi\jpsi$ mass near 4260\mevcc. They are obtained by subtracting the
distribution from neighboring $\pipi\jpsi$ mass regions
(see text). The solid histogram represents ISR Y(4260) Monte Carlo
events.}
\label{fig:mrec}
\end{figure}

\begin{figure}[!h]
\centering
    \includegraphics[height=7cm]{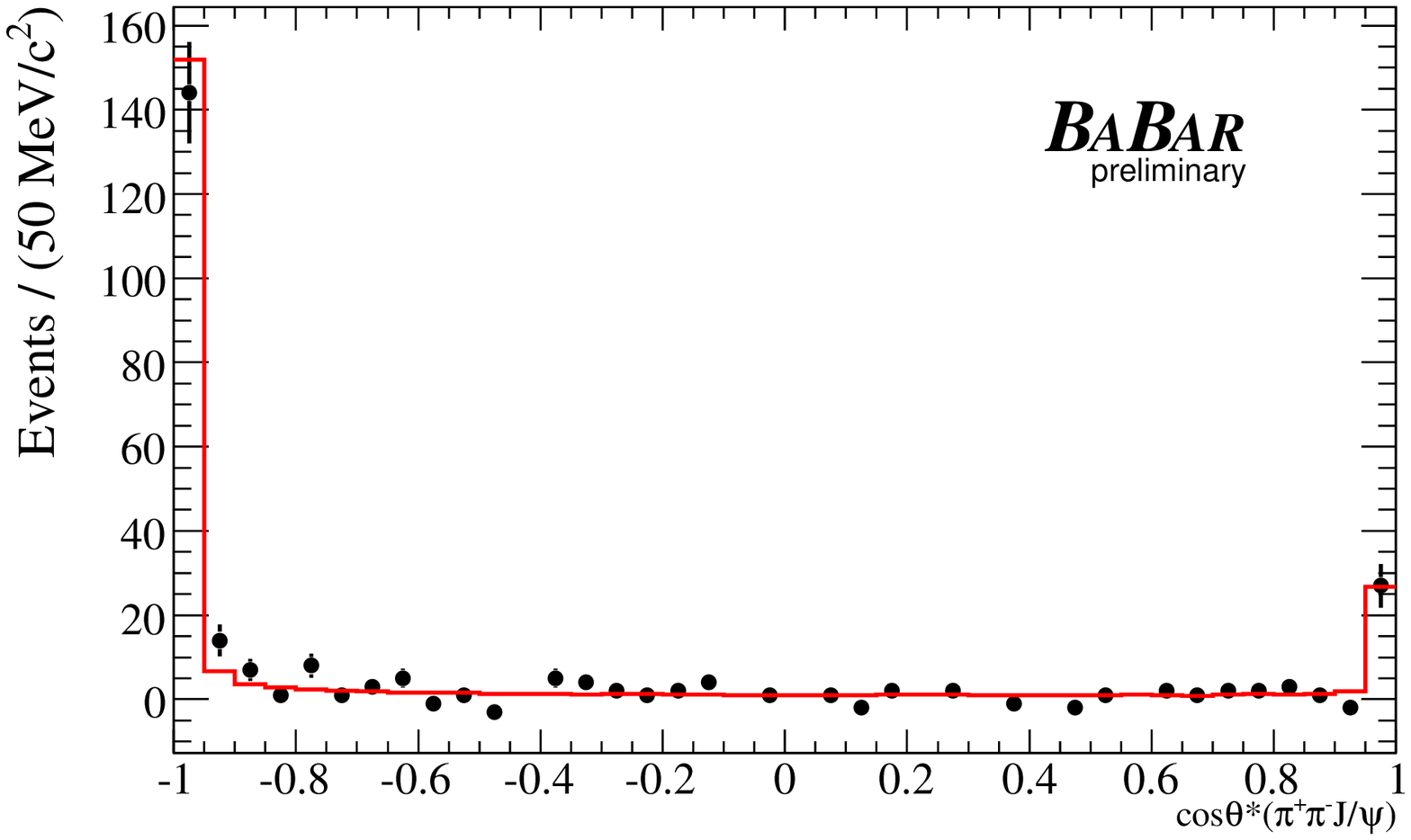}
   \caption{The distribution of $cos\theta_{\pipi\jpsi}^*$. 
   The points represent the
data events passing all selection criteria
and having a $\pipi\jpsi$ mass near 4260\mevcc. They are obtained by subtracting the 
distribution from neighboring $\pipi\jpsi$ mass regions
(see text). The solid histogram represents ISR Y(4260) Monte Carlo
events.}
\label{fig:cos}
\end{figure}

The typical ISR-production signature of the Y(4260) can be obtained
by subtracting distributions for events with $\pipi\jpsi$
mass in the region (3.95,4.1)\gevcc and (4.4,4.55)\gevcc from those with mass
in the signal region defined as (4.1,4.4)\gevcc. The distributions of $m_{rec}^2$ and of the 
cosine of the $\pipi\jpsi$ system polar angle in the c.m. frame are shown in figure \ref{fig:mrec} and \ref{fig:cos}
along with the corresponding distributions for ISR Y(4260)
Monte Carlo events.

\section{RESULTS}
\label{sec:Results}
We have performed an unbinned maximum likelihood fit to the $\pipi\jpsi$ invariant mass distribution between
3.8\gevcc and 5\gevcc. The signal probability density function (PDF) is described by convolving a 
non-relativistic Breit-Wigner function with a Gaussian
and we use a first degree polynomial function to describe the background. 
All the fit parameters for the signal and the background PDF are floating, except the
Gaussian sigma which is fixed to the resolution found at the $\psip$ in data but linearly scaled 
to the Y mass value based on the Monte Carlo simulation.

\begin{figure}[!h]
\centering
    \includegraphics[height=7cm]{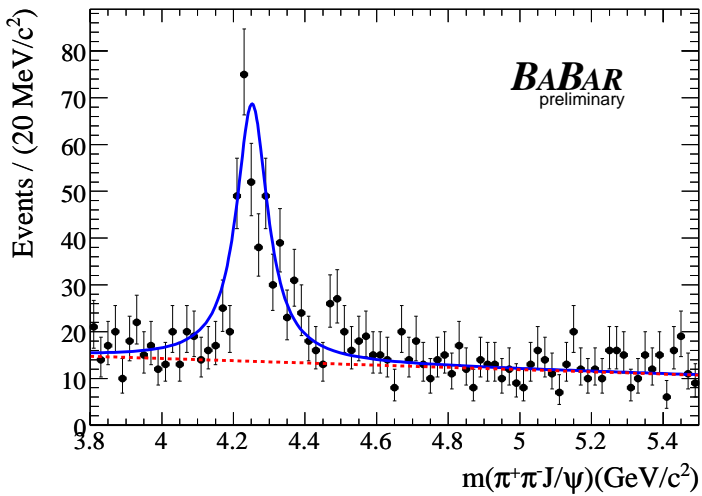}
   \caption{The $\pipi\jpsi$ invariant mass spectrum in the range 3.8-5.5\gevcc.
   The points with error bars represent the selected data, the solid curve shows the
   result of the fit described in the text, while the dashed curve represents the background
   component.}
\label{fig:fit}
\end{figure}

From the fit shown in figure \ref{fig:fit} we obtain 
a signal yield $N_Y = 344 \pm 39$ and resonance mass 
$m_Y = 4252 \pm6~^{+2}_{-3}$\mevcc and width $\Gamma_Y = 105 \pm 18~^{+4}_{-6}$\mevcc. 
Systematic uncertainties include contributions from the fitting procedure 
(evaluated by changing the fit range and the background PDF), the mass scale, the mass-resolution
function and the dependence on the model of Y(4260)$\to\pipi\jpsi$ decay. They have been added in
quadrature.

To measure ${\cal B}(\pipi\jpsi)\Gamma_{e^+e^-}$ we use 
Eq.~\ref{eq:Gee}, where $N(\gamma\,\psip)$, $N(\gamma\,Y)$, $m(\psip)$, $m(Y)$, $\varepsilon(\psip)$, $\varepsilon(Y)$
and $W(\psip)$, $W(Y)$ are the number of events, mass, selection efficiency and the photon emission probability density
function for the $\psip$ and Y(4260) respectively. 
\begin{linenomath}
\begin{multline} \label{eq:Gee}
 \frac{\Gamma_{ee}(Y)\BR(Y\!\to\pipi\jpsi)}
      {\Gamma_{ee}(\psip)\BR(\psip\!\to\pipi\jpsi)}  \\
   = \Bigl( \frac {N(\gamma\,Y)} {N(\gamma\,\psip)} \Bigr) \cdot
     \Bigl( \frac {m(Y)} {m(\psip)}        \Bigr) \cdot
     \Bigl( \frac {\varepsilon(\psip)} {\varepsilon(Y)}  \Bigr) \cdot
     \Bigl( \frac {W(\psip)} {W(Y)}  \Bigr)
\end{multline}
\end{linenomath}
This way, 
the entire uncertainties of integrated luminosity and $\BR(\jpsi\!\to\ell^+\ell^-)$ 
are canceled. Most uncertainties of the selection efficiency, particle ID efficiency, 
tracking efficiency, and photon emission probability density $W(s,x)$ are 
canceled out.
Meanwhile, we also introduce some new uncertainties pertaining to the 
ISR $\psip$ such as $\BR(\psip\!\to\pipi\jpsi)$, $\Gamma_{\psip\!\to\ee}$, and 
statistical uncertainty of the $N(\psip)$.
We obtain ${\cal B}(\pipi\jpsi)\Gamma_{e^+e^-} = (7.5\pm 0.9~\pm0.8)$~eV.

A recent BELLE analysis reported evidence for a broad enhancement 
with mass $m = 4008 \pm 40~ ^{+114}_{-28}$\mevcc and width 
$\Gamma = 226 \pm 44~ \pm 87$\mevcc \cite{:2007sj}. 
We cannot confirm this observation and obtain an upper limit 
${\cal B}(\pipi\jpsi)\Gamma_{e^+e^-} < 0.7$ eV at 90\% C.L. for this state.
 
We also study the $\pipi$ invariant mass distribution for Y(4260) 
events, obtained by dividing the total sample into 
several dipion invariant mass regions and fitting the Y(4260) peak 
in each one of these regions. 

\begin{figure}[!h]
\centering
    \includegraphics[height=7cm]{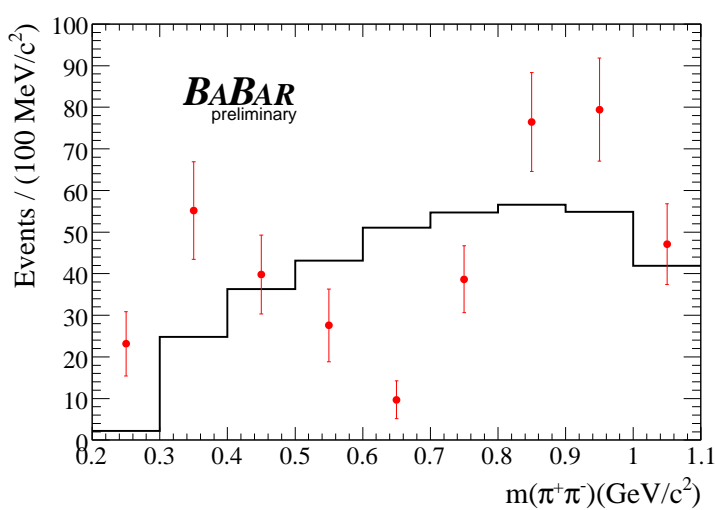}
   \caption{Di-pion invariant mass distribution of Y(4260) events: the points with error
   bars are signal events, the histogram is the phase-space distribution from MC events.}
\label{fig:dipion}
\end{figure}

The results reported in figure \ref{fig:dipion} show that, in Y(4260) decays, the $\pipi$ invariant mass
distribution does not agree with the MC based on phase space and tends to peak
at large values.

\section{CONCLUSIONS}
\label{sec:Conclusions}
In summary, we have analyzed initial-state radiation events to study the process 
$e^+e^-\to\pipi\jpsi$ across the charmonium mass range. We observe $344 \pm 39$ 
Y(4260) events with $m_Y = 4252 \pm 6~^{+2}_{-3}$\mevcc and $\Gamma_Y = 105 \pm 18~^{+4}_{-6}$\mevcc. 
The $\pipi$ invariant mass distribution, in Y(4260) region, tends to peak at large values consistently with
other studies \cite{ct:babar-Y}  \cite{:2007sj}.
There is no evidence for the 
broad enhancement reported by BELLE around 4.05\gevcc \cite{:2007sj} and we obtain an upper limit 
${\cal B}(\pipi\jpsi)\Gamma_{e^+e^-} < 0.7$ eV at 90\% C.L. for this state.

\section{ACKNOWLEDGMENTS}
\label{sec:Acknowledgments}

We are grateful for the 
extraordinary contributions of our \pep2\ colleagues in
achieving the excellent luminosity and machine conditions
that have made this work possible.
The success of this project also relies critically on the 
expertise and dedication of the computing organizations that 
support \babar.
The collaborating institutions wish to thank 
SLAC for its support and the kind hospitality extended to them. 
This work is supported by the
US Department of Energy
and National Science Foundation, the
Natural Sciences and Engineering Research Council (Canada),
the Commissariat \`a l'Energie Atomique and
Institut National de Physique Nucl\'eaire et de Physique des Particules
(France), the
Bundesministerium f\"ur Bildung und Forschung and
Deutsche Forschungsgemeinschaft
(Germany), the
Istituto Nazionale di Fisica Nucleare (Italy),
the Foundation for Fundamental Research on Matter (The Netherlands),
the Research Council of Norway, the
Ministry of Education and Science of the Russian Federation, 
Ministerio de Educaci\'on y Ciencia (Spain), and the
Science and Technology Facilities Council (United Kingdom).
Individuals have received support from 
the Marie-Curie IEF program (European Union) and
the A. P. Sloan Foundation.

\end{document}